\documentclass[runningheads]{llncs}
\usepackage{graphicx}
\usepackage{amsmath}
\usepackage{amssymb}
\usepackage{booktabs}
\usepackage{multirow}
\usepackage{todonotes}
\usepackage{marvosym}

\newcommand\blfootnote[1] 
{
	\begingroup
	
	\footnotetext{#1}
	\addtocounter{footnote}{-1}
	\endgroup
}

\begin{document}
\title{Real-Time Mapping of Tissue Properties for Magnetic Resonance Fingerprinting}
%
%
\author{Yilin Liu\inst{1} 
\and 
Yong Chen\inst{3}\textsuperscript{,\,\Letter} 
\and
Pew-Thian Yap\inst{2,1}\textsuperscript{,\,\Letter}} 
\authorrunning{Liu et al.}
%
\institute{Department of Computer Science, University of North Carolina at Chapel Hill, USA \and
Department of Radiology and Biomedical Research Imaging Center (BRIC), University of North Carolina at Chapel Hill, USA \\
\and Department of Radiology, Case Western Reserve University, Cleveland, USA\\
\email{yxc235@case.edu, ptyap@med.unc.edu}
}
\maketitle              
\begin{abstract}
Magnetic resonance fingerprinting (MRF) is a relatively new multi-parametric quantitative imaging method that involves a two-step process: (i) reconstructing a series of time frames from highly-undersampled non-Cartesian spiral k-space data and (ii) pattern matching using the time frames to infer tissue properties (e.g., $T_1$ and $T_2$ relaxation times). In this paper, we introduce a novel end-to-end deep learning framework to seamlessly map the tissue properties directly from spiral k-space MRF data, thereby avoiding time-consuming processing such as the non-uniform fast Fourier transform (NUFFT) and the dictionary-based fingerprint matching. Our method directly consumes the non-Cartesian $k$-space data, performs adaptive density compensation, and predicts multiple tissue property maps in one forward pass. Experiments on both 2D and 3D MRF data demonstrate that quantification accuracy comparable to state-of-the-art methods can be accomplished within $0.5$ second, which is 1,100 to 7,700 times faster than the original MRF framework. The proposed method is thus promising for facilitating the adoption of MRF in clinical settings.

\blfootnote{ 
	\noindent 
	This work was supported in part by United States National Institutes of Health (NIH) grant EB006733.
}

\end{abstract}
\section{Introduction}
Magnetic resonance fingerprinting (MRF) \cite{ma2013magnetic} is a new quantitative imaging paradigm that allows fast and parallel measurement of multiple tissue properties in a single acquisition, unlike conventional methods that quantify one specific tissue property at a time. MRF randomizes multiple acquisition parameters to generate unique signal evolutions, called ``fingerprints'', that encode information of multiple tissue properties of interest. $1000\sim 3000$ time points are usually acquired and one image is reconstructed for each time point. 
Dictionary matching (DM) is then used to match the fingerprint at each pixel to a pre-defined dictionary of fingerprints associated with a wide range of tissue properties. 
\begin{figure}[t]
\centering
\includegraphics[width=0.8\textwidth]{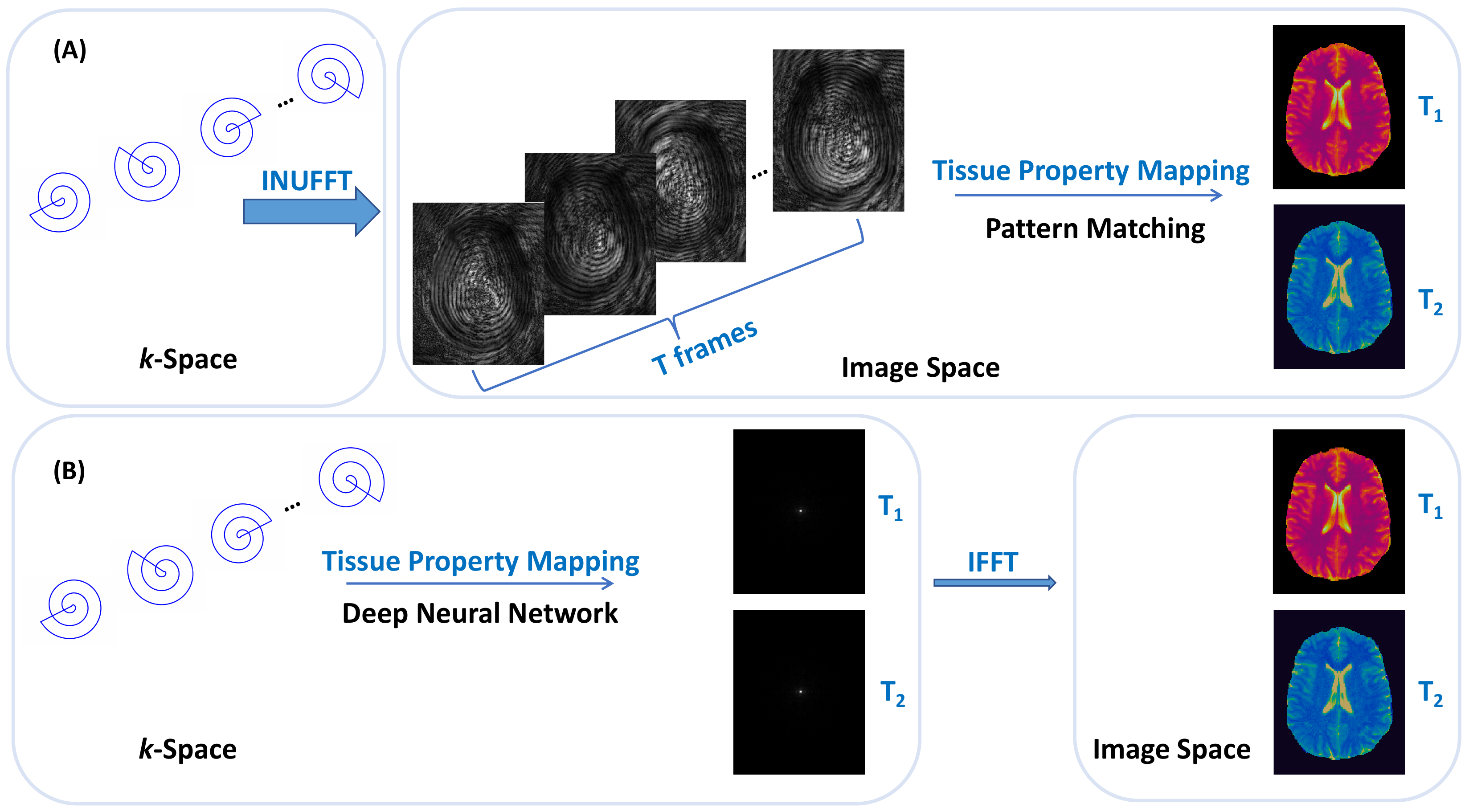}
\caption{(A) The original MRF framework. (B) The proposed framework.}
\end{figure}\label{fig:overview}

To improve the clinical feasibility of MRF, many studies have investigated replacing DM with deep neural networks to accelerate tissue mapping \cite{cohen2018mr,fang2019deep,hoppe2018deep,hoppe2019rinq,fang2020submillimeter}. 
However, these methods, similar to DM, operate on the reconstructed MRF images and are therefore still limited by the speed and computational efficiency of conventional reconstruction methods. 
Particularly, since MRF employs a spiral $k$-space sampling trajectory for robustness to motion \cite{ma2013magnetic}, the reconstruction is non-trivial and more time-consuming than the Cartesian case. 

A major challenge is that the computationally efficient inverse fast Fourier transform (FFT) cannot be directly applied to non-Cartesian data. Besides, the density of the samples varies along the non-Cartesian trajectory and must be compensated for to ensure high-quality reconstruction. Most existing non-Cartesian MRI reconstruction methods thus consist of independent steps that are \emph{not optimized end-to-end}, relying heavily on Non-Uniform Fast Fourier Transform (NUFFT) \cite{fessler2003nonuniform}. 

Fewer deep learning based reconstruction methods focus on non-Cartesian sampling \cite{han2019k,schlemper2019nonuniform} than on Cartesian sampling \cite{zhou2020dudornet,sriram2020grappanet,zhang2019reducing,schlemper2017deep}. AUTOMAP \cite{zhu2018image} attempts to use a fully-connected network (FNN) to learn the full mapping from raw $k$-space data to images, including the Fourier transform (FT). Although FNN makes no assumptions on the sampling pattern and aligns with the nature of FT, the network size is quadratic with the image size $N\times N$ ($O(N^4)$), incurring immense memory costs that limit scalability to large images, especially to high-dimensional MRF data with thousands of time frames. 
Moreover, MRF involves an additional tissue mapping that may require a tailored network architecture based on convolutional neural networks (CNNs) \cite{fang2020submillimeter,fang2019deep} or recurrent neural network (RNNs) \cite{hoppe2019rinq} for optimal performance.

Our aim in this paper is to introduce a framework for real-time tissue quantification directly from non-Cartesian $k$-space MRF data using only regular 2D convolutions and FFT, providing a  computationally more feasible solution to high-dimensional MR data reconstruction and allowing greater flexibility in network design. Mimicking DFT directly with locally-connected CNNs is not effective since every point in the $k$-space has a global effect on the image. Our approach is inspired by the gridding process in NUFFT \cite{bernstein2004handbook,fessler2003nonuniform}. However, instead of explicitly incorporating the memory-costly gridding kernel of NUFFT as in \cite{han2019k,schlemper2019nonuniform}, we show for the first time that gridding and tissue mapping can be performed seamlessly in a single mapping. Experiments on 2D and 3D MRF data demonstrate that our completely end-to-end framework achieves results \emph{on par} with state-of-the-art methods that use more complicated reconstruction schemes while being orders of magnitude faster. To the best of our knowledge, no prior methods have demonstrated the feasibility of end-to-end non-Cartesian MRI reconstruction \emph{for data as high-dimensional as MRF} in a single framework dealing with both reconstruction and tissue mapping simultaneously without the need for NUFFT.  



\section{Methods}

\subsection{Problem Formulation}
With the MRF sequence employed in this study, only $1/48$-th of the full data, i.e., a single spiral, is collected for each time frame for significant acceleration. The original MRF framework first reconstructs an image from each spiral of length $n$ using NUFFT, leading to $T$ highly-aliased images. The image series $\mathbb{C}^{M\times M\times T}$ are then mapped to the corresponding tissue property T1 and T2 maps with image dimensions $M\times M$. 

In contrast, our approach directly maps the highly-undersampled spiral $k$-space MRF data $\mathbb{C}^{n\times T}$ to the Cartesian $k$-space of the $T_1$ or $T_2$ map, and finally to the image space of $T_1$ or $T_2$ map simply via inverse FFT (Fig.~\ref{fig:overview}). 

Let each data point in $k$-space be represented as a location vector $p_i\in\mathbb{R}^2$ and a signal value $f_i\in\mathbb{C}$. To grid the signal $S(q)$, convolution is applied via weighted summation of the signal contributions of $K$ neighboring sampled data points of $q$:
\begin{equation} \label{eq:gridding}
\mathcal{S}(q) = \sum_{i=1}^{K} f_{i}g(p_i - q)d_i,
\end{equation} 
where $g(\cdot)$ denotes the gridding kernel centered at $q$,
and $d_i$ is the density compensation factor for data point $p_i$. Points in sparsely sampled regions are associated with greater compensation factor.
Density compensation is required because in non-Cartesian imaging, the central $k$-space (low-frequency components) is usually more densely sampled than the outer $k$-space (high-frequency components). 

\subsection{Proposed Framework}
Instead of performing gridding in $k$-space and tissue mapping in image space separately as in most existing methods \cite{cohen2018mr,fang2019deep}, we propose to perform tissue mapping  \emph{directly from $k$-space}. This allows gridding and tissue mapping for thousands of time frames to be performed simultaneously via a single CNN, which is key to achieving real-time tissue quantification.
Applying CNNs to non-Cartesian data, however, is not straightforward. Here, without actually interpolating each grid point on the Cartesian $k$-space of MRF frames, our key idea is to directly use the signal time courses of $K$ nearest neighboring spiral points of a target grid point $q$ to infer the corresponding tissue properties (Fig. 2(a)), based on  their relative positions to the target grid point and their densities (Fig. 2(b)). 
 K Nearest-Neighbor (KNN) search and density estimation only need to be computed once for each trajectory and pre-stored; therefore the required time cost is negligible. 
Individual components of the proposed framework are described next.

\begin{figure}[tb]
	\label{fig:archit}
	\centering
	\includegraphics[width=\textwidth]{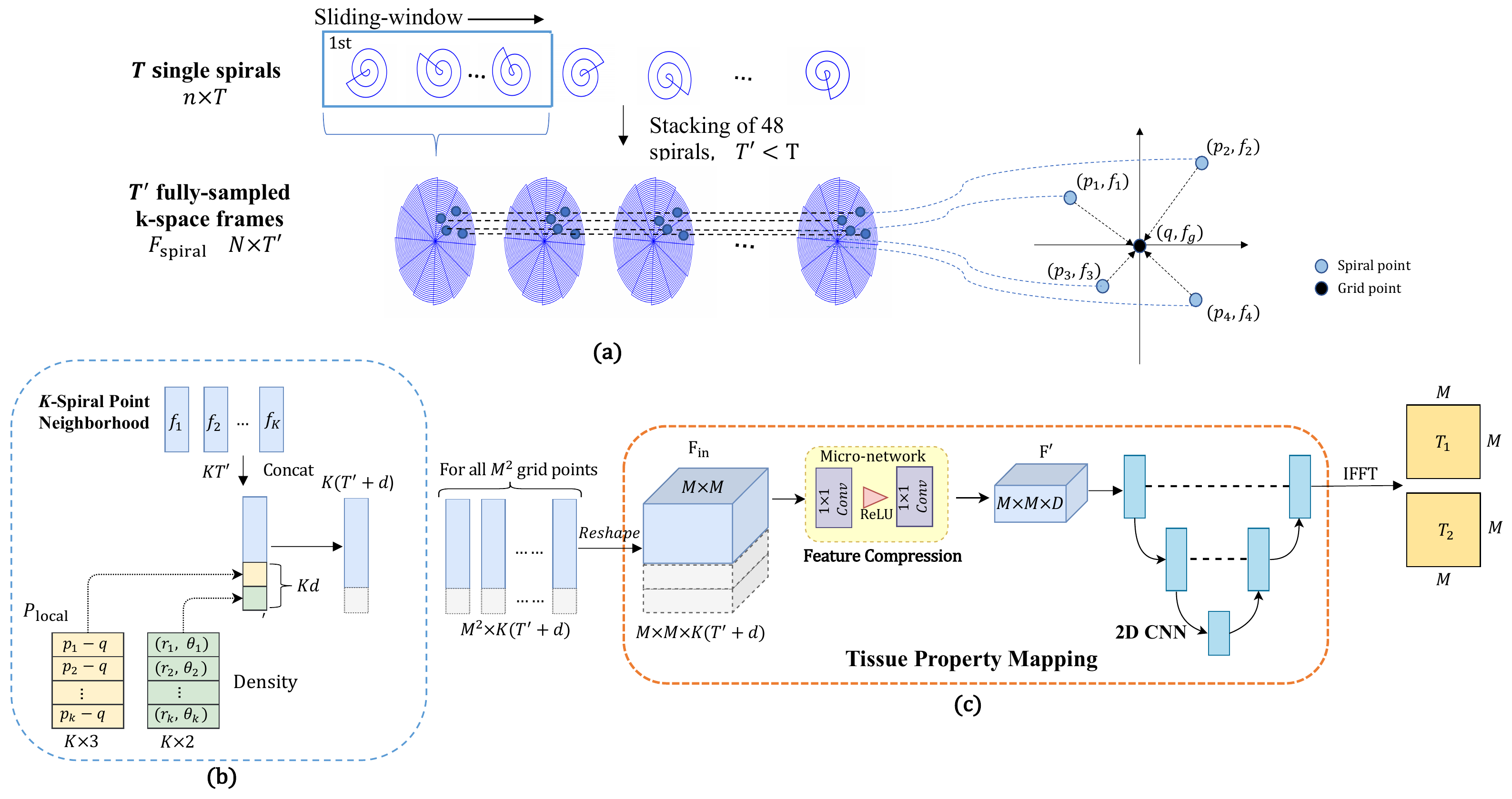}
	\caption{Illustration of the proposed method. The point distribution features ($d=5$) consist of the relative Cartesian offsets of $p_i$ with respect to $q$, the radial distance of $p_i$ with respect to $q$, and the density of $p_i$ represented as the polar coordinates of $p_i$ with respect to the $k$-space center.}
\end{figure} 

\subsection{Sliding-Window Stacking of Spirals}
In MRF, data are typically sampled incoherently in the temporal dimension via a series of rotated spirals. Each MRF time frame is highly undersampled with one spiral. 
Here, we combine every 48 temporally consecutive spirals in a sliding-window fashion for full $k$-space coverage (Fig.~2(a, Left)). This reduces the number of time frames from $T$ to $T'=T/48$ and allows each spiral point to be associated with a $T'$ dimensional feature vector $f_i$. The input to the network thus become: $F_{\text{spiral}}=\{f_1, f_2,...,f_N\}, f_i\in\mathbb{C}^{T'}$. From Eq.~\eqref{eq:gridding}, sampled points are gridded based only on their relative positions to the target grid point, i.e., $p_i-q$. Thus, as exemplified in Fig.~2(a, Right), different $f_i$ contributes differently according to its spatial proximity with respect to the center grid point 
$q$
in a $K$-point local neighborhood.

\subsection{Learned Density Compensation}\label{sec:density}
In non-Cartesian imaging, measurement density varies in $k$-space and is typically dense at the center and sparse at the peripheral of $k$-space. Density compensation (DC) can thus be viewed as a function of data location on a $k$-space sampling trajectory with respect to the $k$-space center.
This is different from gridding using local weighting with respect to a target grid point. Thus, we propose to parameterize the DC function using 2D polar coordinates of the sampled points:
\begin{equation}\label{eq:polar}
\mathnormal{d_i = f_{\text{dc}}(r_i,\theta_i)},
\end{equation}
where $r_i\geq 0$ and $0\leq\theta_i< 2\pi$. 
Straightforward choices of $f_{\text{dc}}$ are $d_i = \frac{\theta_i}{r_i}$ and $d_i = r_i\theta_i$. However, rather than fixing and handcrafting $f_{dc}$, we learn the DC function to adapt to different sampling trajectories via a network that is sensitive to sample locations. This is achieved by directly concatenating the polar coordinates with the spiral point features, inspired by ``CoordConv'' \cite{liu2018intriguing}. By simply giving the convolution access to the input coordinates, the network can adaptively decide where to compensate by learning different weights for the density features associated with different spiral points. This is unlike conventional methods where DC weighting functions are computed analytically \cite{hoge1997density} or iteratively \cite{pipe1999sampling}. See \cite{liu2018intriguing} for more information on  translation-variant CoordConv.

\subsection{Tissue Mapping via Agglomerated Neighboring Features}\label{sec:AgglomeratingFeatures} 
The features for each target grid point are agglomerated from its $K$ nearest neighbors from a stack of $T'$ spirals. This transforms the spiral data $F_{\text{spiral}}$ to a grid, allowing regular 2D convolutions to be applied directly. 
Concatenating point features with additional $d$-dimensional point distribution information required by gridding and density compensation leads to input $F_{\text{in}}\in\mathbb{R}^{M\times M\times K(T'+d)}$. Since our framework does not emphasize on and is not limited to a certain network architecture, we extend an existing U-Net \cite{ronneberger2015u} based MRF tissue quantification network \cite{fang2019deep} to make it fully end-to-end, mapping the agglomerated features $F_{\text{in}}$ directly to the corresponding tissue property maps. 
To improve computational efficiency, a micro-network is employed preceding the quantification network to reduce the dimensionality of each target grid feature vector $f\in\mathbb{R}^{K(T'+d)}$ by a shared linear transformation $W\in \mathbb{R}^{K(T'+d)\times D}$, implemented as an $1\times 1$ convolution: 
\begin{equation} 
f_{j}' = \text{ReLU}(\text{Conv}([f_1;f_2;...;f_K])),
\end{equation} 
where $[\cdot;\cdot]$ denotes concatenation, $f_j'\in \mathbb{R}^{D}$, and $j \in \{1,2,\hdots, M\times M\}$. 
The resulting feature map $F'\in \mathbb{R}^{M\times M \times D}$ is then fed to the quantification network. 


\subsubsection{Network Parameters and Training.}
Our network backbone consists of a micro-network and a 2D U-Net, which is $\sim 10^5$ lighter than AUTOMAP \cite{zhu2018image}. AUTOMAP is computationally expensive when applied to MRF ($\sim 5\times 10^{12}$ params). The micro-network is composed of four $1\times 1$ convolutional layers, each followed by batch normalization and ReLU. The number of output channels of all $1\times 1$ convolutions is $D$ ($D=64$ for T1, and $D=164$ for T2, chosen by cross validation). The input channel number of the micro-network is $K(T'+d)$, where $K=4$. The network was trained in batches of 2 samples and optimized via ADAM with an initial learning rate of $0.0002$, which was decayed by $99\%$ after each epoch. Following \cite{fang2019deep}, relative-L1 was used as the objective function. Two GPUs (TITAN X, $12$G) were used for training.
 




\section{Experiments and Results}
\subsubsection{Datasets.} 2D MRF datasets were acquired from six normal subjects, each consisting of $10$ to $12$ scans. For each scan, a total of $2,304$ MRF time points were acquired and each contains only one spiral readout of length $n=2,052$. 
Two 3D MRF datasets were used for evaluation. The first 3D MRF dataset with a spatial resolution of $1.2\times 1.2\times 3$ mm$^3$ were collected from three subjects, each covering $64$ slices. A total of $576$ time points were acquired for each scan. 
The second 3D MRF datasets were acquired from six volunteers with a high isotropic resolution of $1$\,mm, each covering $144$ slices. $768$ time points were collected for each scan. For both 3D datasets, FFT was first applied in the slice-encoding direction, and then the data of each subject were processed slice-by-slice, just as in the 2D case. All MRI measurements were performed on a Siemens 3T scanner with a $32$-channel head coil. Real and imaginary parts of the complex-valued MRF signals are concatenated. For acceleration, only the first $25\%$ time frames in each 2D MRF scan and the first $50\%$ in each 3D MRF scan were used for training. The training data size for \emph{each} 2D and 3D scan is $32\times 2,052\times 576$ and $32\times 2\rm{,}052\times 288$ (or $384$) (\#~coils$\times$ \#~spiral readouts $\times$ \#~time frames), respectively.
The ground-truth $T_1$ and $T_2$ maps with $256\times 256$ voxels were obtained via dictionary matching using all $2\rm{,}304$ time frames in 2D MRF and all $576$ (or $768$) time frames in 3D MRF.

\subsubsection{Experimental Setup.}
1) We compared our end-to-end approach with four state-of-the-art MRF methods: a U-Net based deep learning method (\textbf{SCQ}) \cite{fang2019deep}, dictionary matching (\textbf{DM}) \cite{ma2013magnetic}, SVD-compressed dictionary matching (\textbf{SDM})  \cite{mcgivney2014svd}, and a low-rank approximation method (\textbf{Low-Rank}) \cite{ma2017applications}. Note that these competing methods require first reconstructing the image for each time frame using NUFFT \cite{fessler2003nonuniform}. 
Leave-one-out cross validation was employed. 2) We also compared our adaptive gridding with typical handcrafted gridding methods, and investigated the effects of including the relative positions and density features. 3) As a proof of concept, we applied our method on the additional high-resolution 3D MRF dataset for qualitative evaluation.

\begin{table*}[hbt!]
\centering
\caption{Quantitative comparison on the 2D MRF dataset with 4$\times$ under-sampling. MAE is computed relative to the ground truth (unit: \%). Times reported for reconstruction and pattern matching are per-slice averages.}\label{ta:2Dcomp}
\begin{small}
\centering
\begin{tabular}{@{}lccllccccc@{}}
\toprule
Method   & \multicolumn{2}{c}{MAE}       & \multicolumn{2}{c}{SSIM}                        & \multicolumn{2}{c}{NRMSE}                              & Recon. (s) & Patt. Match. (s) & Total (s)       \\ \midrule
         & T1            & T2            & \multicolumn{1}{c}{T1} & \multicolumn{1}{c}{T2} & T1                                  & T2              & \multicolumn{1}{l}{}    & \multicolumn{1}{l}{}      & \multicolumn{1}{l}{} \\
DM       & \textbf{2.42} & 10.06         & \textbf{0.994}         & 0.954                  & \multicolumn{1}{l}{\textbf{0.0150}} & 0.0421          & 467                     & 25                        & 492                  \\
SDM      & 2.42          & 10.05         & 0.994         & 0.954                  & 0.0150                              & 0.0421          & 467                     & 10                        & 477                  \\
Low-Rank & 2.87          & 8.17          & 0.991                  & 0.960                  & 0.0156                              & \textbf{0.0302} & 3133                    & 25                        & 3158                 \\
SCQ      & 4.87          & 7.53          & 0.992                  & 0.968                  & 0.0217                              & 0.0309          & 9.73                    & 0.12                      & 9.85                \\ \midrule
Ours     & 4.24          & \textbf{7.09} & 0.986                  & \textbf{0.972}         & 0.0258                              & 0.0335          & --                      & --                     & \textbf{0.41}                 \\ \bottomrule
\end{tabular}
\end{small}
\end{table*}

\begin{figure*}[h!]
   \centering
  \includegraphics[width=\textwidth]{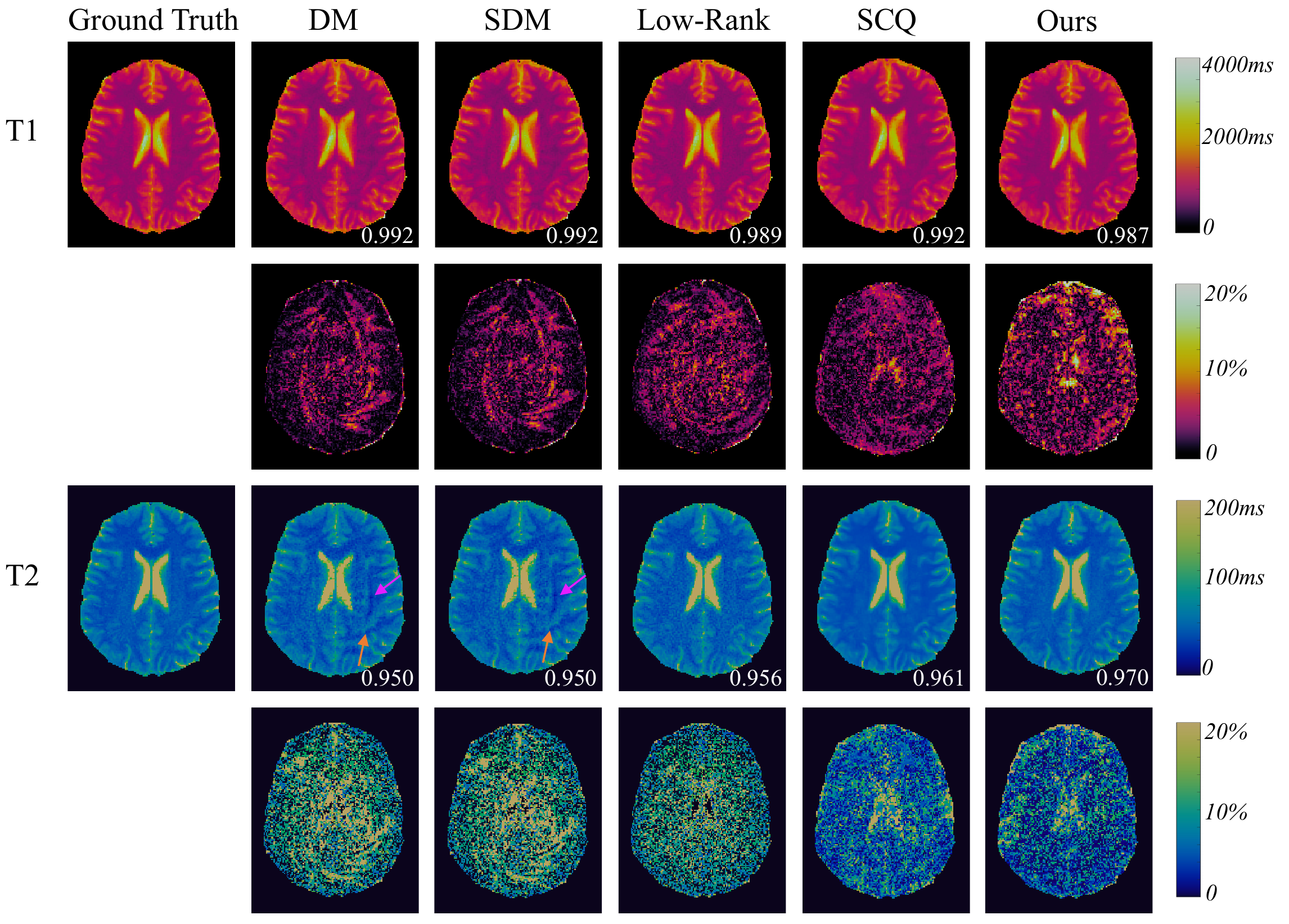}
  \caption{Example 2D MRF results and the associated error maps with 4$\times$ under-sampling. Artifacts are indicated by arrows. SSIM is reported at the bottom right.}
  \label{fig:2DComp}
\end{figure*}

\subsubsection{Results and Discussion.} 
\begin{figure*}[t]
   \centering
  \includegraphics[width=\textwidth]{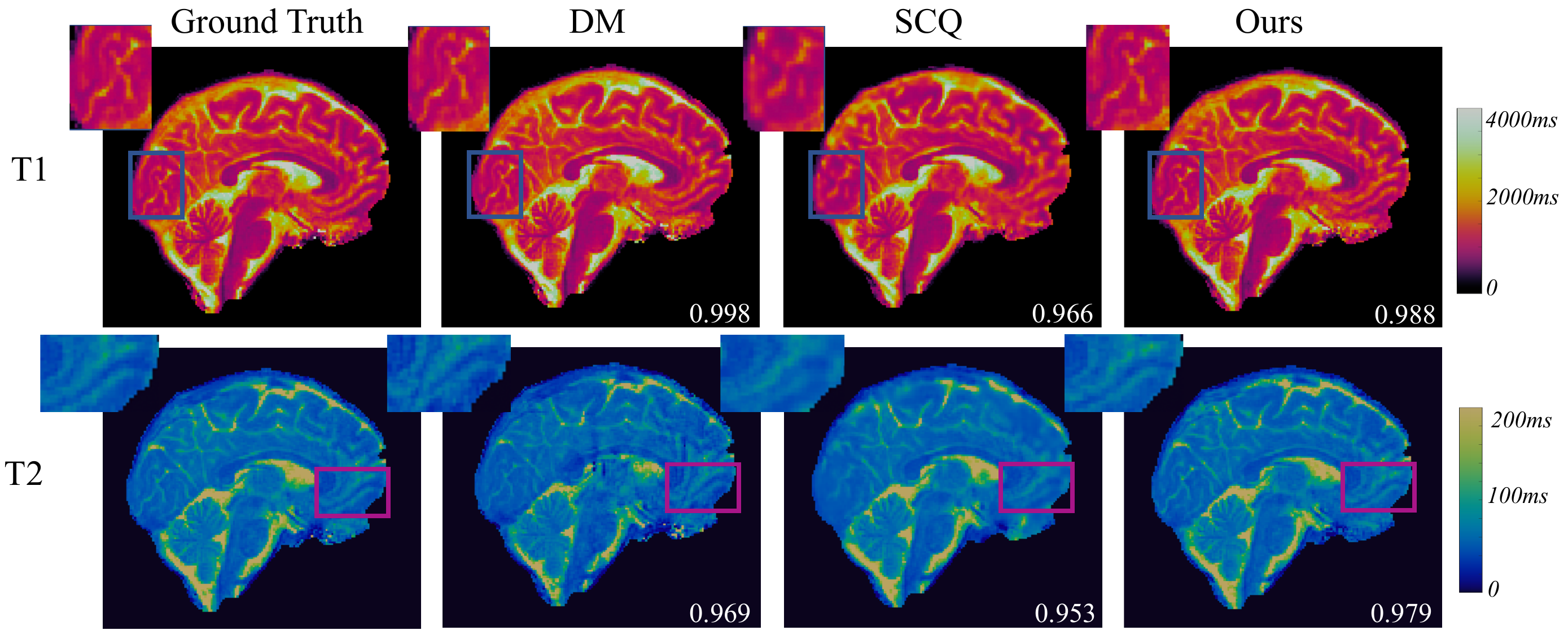}
  \caption{Example 3D MRF results with 2$\times$ under-sampling and $1.2\times 1.2\times 3$\,mm$^3$ resolution. SSIM is reported at the bottom right.}
  \label{fig:3DComp}
\end{figure*}

\begin{figure*}[t]
   \centering
  \includegraphics[width=\textwidth]{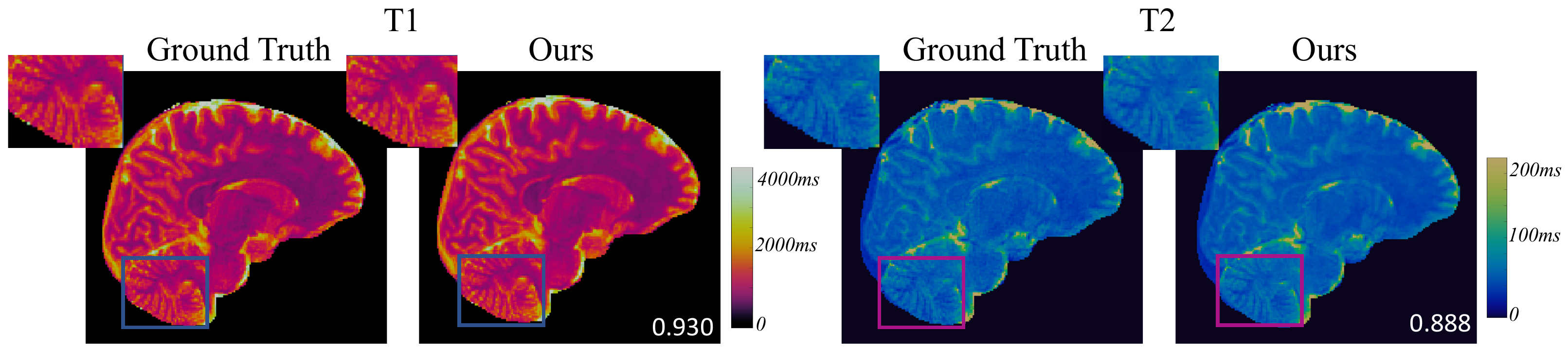}
  \caption{Example high-resolution 3D MRF results with 2$\times$ under-sampling and $1$ mm isotropic resolution. SSIM is reported at the bottom right.}
  \label{fig:highres_3D}
\end{figure*}

As shown in Table~\ref{ta:2Dcomp} and Table~\ref{ta:3Dcomp}, our method performs overall best in T2 quantification accuracy and achieves competitive accuracy in T1 quantification with processing speed 24 times faster than a CNN method and 1,100 to 7,700 times faster than DM methods. Particularly, for 3D MRF, our method performs best for most metrics. Qualitative results are shown in Fig.~\ref{fig:2DComp} and Fig.~\ref{fig:3DComp}. The higher T1 than T2 quantification accuracy is consistent with previous findings \cite{zhao2016maximum,fang2019deep}. Due to the sequence used in this study, the early portion of the MRF time frames, which were used for training, contain more information on T1 than T2. Hence, all methods are more accurate in T1 quantification. DM methods exhibit significant artifacts in T2 as indicated by the arrows in Fig.~\ref{fig:2DComp}. Representative results for the additional high-resolution 3D MRF data are shown in Fig.~\ref{fig:highres_3D}. In the ablation study shown in Table~\ref{ta:3}, our adaptive gridding performs better than  typical handcrafted gridding methods.


\begin{table*}[t]
\centering
\caption{Quantitative comparison using the first 3D MRF data with 2$\times$ under-sampling and a resolution of $1.2\times 1.2\times 3$\,mm$^3$. MAE is computed relative to the ground truth (unit: \%). Times reported for reconstruction and pattern matching are per-slice averages.}\label{ta:3Dcomp}
\begin{small}
\centering
\begin{tabular}{@{}ccclcclcllccc@{}}
\toprule
Method               & \multicolumn{2}{c}{MAE}        &  & \multicolumn{2}{c}{SSIM}                &  & \multicolumn{2}{c}{NRMSE}                             &  & \multicolumn{1}{l}{Recon. (s)} & \multicolumn{1}{l}{Patt. Match. (s)} & \multicolumn{1}{l}{Total (s)} \\ \midrule
\multicolumn{1}{l}{} & T1            & T2             &  & T1             & \multicolumn{1}{l}{T2} &  & T1              & T2                                  &  & \multicolumn{1}{l}{}           & \multicolumn{1}{l}{}                 & \multicolumn{1}{l}{}          \\
DM                   & \textbf{5.89} & 12.19          &  & \textbf{0.996} & \textbf{0.968}         &  & 0.0415          & \multicolumn{1}{c}{0.0521}          &  & 140.56                         & 17.01                                & 157.57                        \\
SCQ                  & 16.58         & 16.74          &  & 0.933          & 0.919                  &  & 0.0652          & 0.0479                              &  & 8.58                           & 0.11                                 & 8.69                          \\
Ours                 & 9.14          & \textbf{11.78} &  & 0.980          & \textbf{0.968}         &  & \textbf{0.0389} & \multicolumn{1}{c}{\textbf{0.0323}} &  & --                             & --                                   & \textbf{0.33}                 \\ \bottomrule
\end{tabular}
\end{small}
\end{table*}

\begin{table}[t]
\caption{Comparison of our adaptive gridding method with typical handcrafted gridding methods, and effects of including relative positions and density features.}\label{ta:3}
\centering

\begin{tabular}{@{}lc|ccc|clcc@{}}
\toprule
\multicolumn{2}{l|}{\multirow{2}{*}{\begin{tabular}[c]{@{}l@{}}Gridding \\ Method\end{tabular}}} & \multirow{2}{*}{Average} & \multirow{2}{*}{Bilinear} & \multirow{2}{*}{Gaussian} & \multicolumn{4}{c}{Ours}                                        \\ \cmidrule(l){6-9} 
\multicolumn{2}{l|}{}                                                                          &                          &                           &                           & No xy/density & \multicolumn{1}{c}{xy} & density & xy+density    \\ \midrule
\multicolumn{1}{c}{\multirow{2}{*}{\begin{tabular}[c]{@{}c@{}}MAE\\ (\%)\end{tabular}}}  & T1  & 5.59                     & 5.27                      & 5.53                      & 5.24          & 4.34                   & 4.48    & \textbf{4.24} \\
\multicolumn{1}{c}{}                                                                     & T2  & 7.74                     & 8.48                      & 7.95                      & 9.05          & 8.43                   & 7.37    & \textbf{7.09} \\ \bottomrule
\end{tabular}

\end{table} 



\section{Conclusion}
In this paper, we introduced a novel and scalable end-to-end framework for direct tissue quantification from non-Cartesian MRF data in milliseconds. With $0.5$s per slice, $120$ slices for whole-brain coverage can be processed in one minute, allowing timely re-scan decisions to be made in clinical settings without having to reschedule additional patient visits. It should be noted that the U-Net based network backbone can be replaced with a more advanced architecture to further boost quantification accuracy. Our framework is also agnostic to the data sampling pattern, and thus can be potentially adapted to facilitate other non-Cartesian MRI reconstruction tasks. 
We believe that our work will improve the clinical feasibility of MRF, and spur the development of fast, accurate and robust reconstruction techniques for non-Cartesian MRI.
 \bibliographystyle{splncs04}
 \bibliography{ref}
%






\end{document}